\documentclass[aps,groupedaddress,prb,reprint,showpacs]{revtex4-1}
\usepackage{graphicx}
\usepackage{amssymb}
\begin{document}

\title{Role of covalent and metallic intercalation on the electronic properties of epitaxial graphene on SiC(0001)}

\author{I. Deretzis}
\email{ioannis.deretzis@imm.cnr.it}
\affiliation{Istituto per la Microelettronica e Microsistemi (CNR-IMM), VIII strada 5, I-95121 Catania, Italy}

\author{A. La Magna}
\affiliation{Istituto per la Microelettronica e Microsistemi (CNR-IMM), VIII strada 5, I-95121 Catania, Italy}
\date{\today}

\begin{abstract}
We present an orbital-resolved density functional theory study on the electronic properties of hydrogen and lithium intercalated graphene grown on the Si face of SiC. Starting from the $(6\sqrt3\times6\sqrt3)R30^{\circ}$ surface reconstruction of the graphene/SiC heterosystem, we find that both H and Li can restore the ideal structural characteristics of the two nonequivalent junction parts (i.e. graphene and the SiC substrate) when inserted at the interface. However, the chemical/electrostatic interactions remain different for the two cases. Hence, H-intercalated epitaxial graphene is subject to a sublattice symmetry-breaking electronic interference that perturbs the Dirac point, whereas Li intercalation gives rise to a highly $n$-doped system due to a nonuniform delocalization of Li charges. Results bring to discussion the role of substrate engineering in epitaxial graphene on SiC.
\end{abstract}

\maketitle


\section{Introduction}
\label{intro}

Graphene is a single-atom-thick crystal of $sp^2$-hybridized carbons that exhibits exceptional electronic properties due to its linear dispersion relation around the Dirac point \cite{2009RvMP...81..109C}. During the last years the quest for its use in fundamentals and applications has boosted the research on processes that could allow for a controlled synthesis of ordered graphene layers. Within this context, epitaxial growth on SiC substrates has emerged as one of the principal technologies for large-scale graphene fabrication \cite{2009NatMa...8..203E,2010JVSTB..28..985D}. The process could be better described as a ``de-growth'' one, since SiC crystals are heated up to temperatures that allow for an extremely higher sublimation rate of the Si atoms with respect to the C ones. The remaining C surface atoms recompose to form thin graphite films. A proper calibration of the process parameters \cite{2009NanoL...9.2605J} can allow for the formation of single graphene layers directly grown on a semi-insulating substrate, i.e. with no need to be transfered elsewhere after the end of the thermal treatment.

Epitaxial graphene can grow on both the Si and C faces of the (0001) surface of hexagonal SiC polytypes \cite{2008JPCM...20F3202H}, or similarly at the (111) surface of cubic SiC \cite{2010ApPhL..96s1910O}. Even if both theory \cite{2010PhRvB..82l1416P,2011ApPhL..98b3113D} and experiment \cite{2009ApPhL..95v3108W} argue that C face epitaxial graphene should electrically suffer less interferences from the substrate, Si face growth results slower \cite{2009NanoL...9.2605J,2008JPCM...20F3202H}. Hence, a better control can be achieved on the formation of single or few-layer structures. In this case a first carbon-rich interface layer with a $(6\sqrt3\times6\sqrt3)R30^{\circ}$ surface reconstruction (often called the buffer layer) strongly binds to the substrate and is the precursor of the overlying graphene films \cite{2008JPCM...20F3202H,2008PhRvL.100q6802K}. Experimental evidence shows that the presence of this carbon layer has a negative influence in the conduction properties of the heterosystem with respect to the $SiO_2$-deposited case: measurements report an enhanced surface polar phonon scattering mechanism \cite{2010ApPhL..97m2101S}, reduced mean-free paths \cite{2010ApPhL..97m2101S} and temperature-dependent mobilities \cite{2011arXiv1103.3997S}, which are characteristic of the diffusive transport regime. To this end, intercalation techniques with various elements have been proposed in the literature \cite{2009PhRvL.103x6804R,2010PhRvB..82t5402V,2011ApPhL..98r4102W,2010PhRvB..81w5408G,2011PhRvB..84l5423E} in order to detach the buffer layer from the substrate and minimize the interface interaction. A general characteristic of these processes is that the intercalated element can form either covalent (e.g. H or Ge) or metallic (e.g. Li or Au) bonds with the SiC(0001) surface and induce a doping level, whose origin is often unclear. 

In this article we study the structural and electronic properties of H and Li-intercalated epitaxial graphene within a comparative approach of functionalization with adatoms that interact in a covalent and a metallic way with the substrate. As a starting point we recognize the problems related to the presence of the buffer layer within a density functional theory (DFT) description of the $(6\sqrt3\times6\sqrt3)R30^{\circ}$ reconstruction (Sec. \ref{interface}). We then proceed with an analysis of the structural and electronic symmetry of the intercalated systems (Sec. \ref{hydrogen} and \ref{lithium} for H and Li respectively), arguing that the resulting chemically modified interfaces are different in terms of effective doping, bonding and electrostatic interactions. Finally, in Sec. \ref{discussion} we discuss  the versatility of substrate engineering in tailoring the physical properties of epitaxial graphene on SiC.

\section{Methodology}
\label{meth}
    
We use the DFT SIESTA computational code \cite{2002JPCM...14.2745S} to perform \textit{ab initio} calculations, treating the electronic correlations within the local density approximation (LDA) \cite{1981PhRvB..23.5048P}, which has been shown to describe properly the experimental aspects of the graphene/SiC interface \cite{2008PhRvL.100q6802K}. Moreover, taking into account the absence of nonlocal dispersive interactions in the LDA, we also perform a comparative analysis for the intercalated systems using different exchange-correlation functionals: in the spirit of Ref. \citenum{2011PhRvB..84c5442J}, apart from the LDA we use the Perdew-Burke-Ernzerhof implementation \cite{1996PhRvL..77.3865P} of the generalized gradient approximation (GGA) and the van der Waals (vdW) functional of Dion \textit{et al} \cite{2004PhRvL..92x6401D,2009PhRvL.103i6102R}. The epitaxial graphene structures comprise of two bilayers of a $(6\sqrt3\times6\sqrt3)R30^{\circ}$ SiC substrate (passivated with H at the bottom of the slab), over which a single $(13 \times 13)$ graphene supercell relaxes in order to satisfy lattice commensuration. The aspect of the correct surface reconstruction is fundamental in the computational modeling of the designated systems. Indeed, models based on the simpler $(\sqrt3\times\sqrt3)R30^{\circ}$ reconstruction \cite{2009ApPhL..95f3111D,2007PhRvL..99g6802M,2007PhRvL..99l6805V}, albeit efficient in describing some of the basic electronic properties of the SiC/graphene interface (e.g. $n$-type doping and presence of the carbon-rich layer), are characterized by: (a) the lack of commensuration for the graphene and the SiC lattices that leads to an 8\% stretching of the graphene sheet, and (b) a small periodically-repeated $(2 \times 2)$ graphene supercell. Repercussions of these aspects with respect to the results obtained by the actual $(6\sqrt3\times6\sqrt3)R30^{\circ}$ surface symmetry are discussed in Sec. \ref{hydrogen}.  

We model three different systems, i.e. the graphene/SiC interface before and after intercalation with H and Li, considering a full and uniform coverage of the substrate by the intercalating element in the form of a monolayer. A basis set of double-$\zeta$ Sankey-type valence orbitals has been used for C, Si and H, while polarization orbitals have been added in the case of Li. Convergence tests on a $(\sqrt3\times\sqrt3)R30^{\circ}$-based system have shown that the previous set can capture the main band structure features obtained by a more accurate basis that includes polarization orbitals for all elements. We exploit the localized character of the atomic orbitals in order to calculate the single-orbital contributions in the composite electronic band structure: for each nonequivalent $k$-point sampled along the closed $\Gamma \rightarrow M \rightarrow K \rightarrow \Gamma$ Brillouin-zone path we calculate the wave functions $\Psi_{n,\mathbf{k}}$ corresponding to the $n$ eigenstates. Based on the linear combination of atomic orbitals $\Psi_{n,\mathbf{k}} = \sum_i c_{i} \phi_i$ (where $c_i$ are the weighting coefficients and $\phi_i$ the pseudoatomic orbitals) we calculate the square modulus of the total weight of the single orbitals at the $n^{th}$ eigenstate as: $ w_{n,\mathbf{k}} = \sum_{j} \left| c_{j} \right|^2,$ where the integration runs over those indexes $j \le i$ for which we want to calculate the contributions in the electronic structure. Contrary to the variational approach in the calculation of the valence electronic properties, the electronic contribution of the ionic cores is statically described with norm-conserving Troulier-Martins pseudopotentials \cite{1991PhRvB..43.1993T} that have been tested to accurately reproduce the band structure of hexagonal SiC polytypes and graphene. The minimization of the electron density is achieved by sampling the Brillouin zone with a single-point Monkhorst-Pack grid for the $(6\sqrt3\times6\sqrt3)R30^{\circ}$ model and a $7 \times 7 \times 1$ grid for the $(\sqrt3\times\sqrt3)R30^{\circ}$ one. A mesh cutoff energy of 350 $Ryd$ has been imposed for real-space integration, while all structures have been relaxed with a force criterion of 0.06 eV/\AA{}.

\section{The graphene/SiC(0001) interface}
\label{interface}

\begin{figure}
	\centering
		\includegraphics[width=0.9\columnwidth]{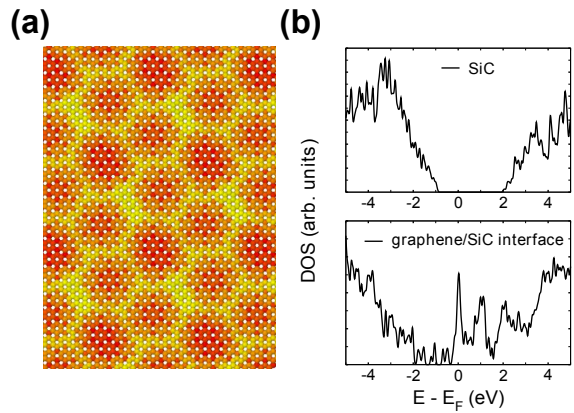}
	\caption{(Color online) (a) Color map topmost view of the $(6\sqrt3\times6\sqrt3)R30^{\circ}$-reconstructed first carbon-rich layer of an epitaxial graphene system showing the vertical positions of the C atoms. Here, a gradual yellow (bright in grayscale) to red (dark in grayscale) coloring indicates bigger  to smaller distances from the substrate. (b) Density of states as a function of energy for an ideal 4H-SiC crystal (upper) and the graphene/SiC(0001) interface (lower).}
	\label{fig:buffer}
\end{figure}

As a first step towards understanding the need for intercalation in epitaxial graphene systems grown on the Si face of SiC, we calculate the structural and electronic properties of the $(6\sqrt3\times6\sqrt3)R30^{\circ}$ graphene/SiC interface. The relaxation of the  corresponding supercell gives rise to a corrugated carbon layer with a thickness of $\sim$ 1.5 \AA{}. The surface is characterized by height-varying Moir\'e patterns that form hexagons with edges of 10-12 \AA{}, which correspond to the regions of the carbon layer that are more distant from the substrate [see the yellow (bright in grayscale) areas in Fig. \ref{fig:buffer}(a)]. This structural deformation with respect to the ideal lattice topology is also reflected in terms of the chemical bonding, where an interplay between $sp^2$ and $sp^3$ interface interactions appears. Quantitatively, there exists a strong preferential disposition towards the $sp^3$ C-Si covalent bonding, which bounds the carbon layer strongly to the substrate. This aspect comes in contrast with the respective interface of C face epitaxial graphene \cite{2011ApPhL..98b3113D} where interface bonding is inherently weaker, and reflects the weakness of the $\pi$-bond in Si with respect to C. The Si atoms of the (0001) surface that do not covalently bind with the substrate maintain the characteristics of the polar SiC(0001) surface with a pronounced inward relaxation with respect to the ideal position, while they introduce dangling-bond bands within the SiC bandgap\cite{1997PSSBR.202..421P}. This last feature characterizes the electronic properties of the entire graphene/SiC heterostructure, where a half-filled peak in the density of states at the Fermi level ($E_F$) appears [Fig. \ref{fig:buffer}(b)-lower]. Moreover the $2p_z$ orbitals of the carbon epilayer, which originate from C atoms that are not bound to the substrate, give rise to surface states within the SiC bandgap. The key point that arises from these simulations is that the graphene/SiC interface is electrically active. Such feature should have an adverse impact on the device potential of Si face epitaxial graphene systems, since it could be directly related to the manifestation of leakage interface currents during conduction.

\section{Hydrogen intercalation}
\label{hydrogen}

\begin{figure}
	\centering
		\includegraphics[width=\columnwidth]{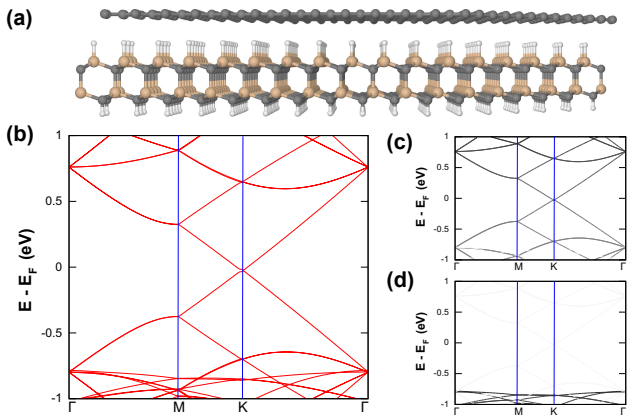}
	\caption{(Color online) (a) Geometry of the relaxed H-intercalated epitaxial graphene system. (b) Total band structure of the H-intercalated system. (c) Contributions of the $2p_z$ orbitals of the graphene layer in the total band structure. (d) Contributions of the substrate orbitals in the total band structure. The z-axis is perpendicular to the graphene plain.}
	\label{fig:hydrogen}
\end{figure}

\begin{table}
	\centering
		\begin{tabular*}{\columnwidth}{@{\extracolsep{\fill}} l c c c }
		\hline 
		\hline 
		model & LDA & GGA & vdW \\
		\hline
	$(\sqrt3\times\sqrt3)R30^{\circ}$ & 130 meV & 80 meV & 15 meV \\
	$(6\sqrt3\times6\sqrt3)R30^{\circ}$ & 8 meV & 5 meV &  5 meV	\\
	\hline 
		\hline
		\end{tabular*}
	\caption{Comparison of the energy gap values ($E_g$) of the H-intercalated epitaxial graphene system for the $(\sqrt3\times\sqrt3)R30^{\circ}$ and the $(6\sqrt3\times6\sqrt3)R30^{\circ}$ models within the LDA, GGA and vdW exchange-correlation functionals.}
	\label{tab:Gaps}
\end{table}

Si-face grown epitaxial graphene can be used as a reference system for the study of a number of fundamental physical properties \cite{2010NatNa...5..186T} without the need of further processing. However, the electrical use of such material within a device concept dictates the need of an interface passivation with functional adatoms through a post-growth chemical process. The simplest example of intercalation from a monovalent element is that of hydrogen, as proposed by Riedl \textit{et al.} \cite{2009PhRvL.103x6804R} and followed by others \cite{2011arXiv1103.3997S,2010SurSc.604L...4V}. Here, starting from the $(6\sqrt3\times6\sqrt3)R30^{\circ}$ reconstruction of the carbon-rich layer, we model H intercalation by assigning a single H atom on top of each Si atom of the substrate's surface. This geometry is experimentally confirmed by infrared absorption spectroscopy \cite{2011arXiv1103.3997S}. Upon relaxation, H atoms covalently bind with the interface Si atoms and restore the ideal $sp^3$ hybridization of the substrate [Fig. \ref{fig:hydrogen}(a)]. As a result, the structurally disordered carbon layer turns into an almost detached flat graphene sheet with a distance of 2.39 \AA{} from the H interface layer and 3.92 \AA{} from the substrate Si atoms. The electronic band structure of this system shows the $(13 \times 13)$ folded graphene bands and confirms the minimal interaction with the substrate (Fig. \ref{fig:hydrogen}(b)). However a small perturbation appears at the Dirac point that opens a bandgap of few millielectronvolts, in analogy with a similar feature calculated for the interaction between graphene and the (0001) surface of SiO$_2$ \cite{2011PhRvL.106j6801N}. We find this perturbation robust for small horizontal shifts of the graphene layer with respect to the substrate. Even if this gap is extremely small, well beyond $k_BT$ at room temperature and difficult to reveal in experiments, its physical origin is of a particular interest. A careful analysis of the electronic Hamiltonian of the composite system excludes an inter-valley contribution on the electronic perturbation. Instead, the origin of this interference can be traced at the breaking of the sublattice symmetry \cite{2007NatMa...6..770Z} due to the unequal coupling between the two interpenetrating triangular lattices of the graphene sheet and the localized Si-H dipoles of the substrate, which have non-equivalent positions with respect to the graphene atoms in the supercell. For the smaller $(\sqrt3\times\sqrt3)R30^{\circ}$ model, this symmetry-breaking effect acquires an extreme form due to the fact that the graphene sheet is defined by only eight periodically reproduced C atoms, and manifests in the form of wider band gaps in the electronic band structure (see Table \ref{tab:Gaps}). In this case, the choice of the exchange-correlation functional becomes critical and both the LDA and the GGA overestimate the graphene-substrate interaction. Contrary, results obtained by the vdW functional indicate that the inclusion of nonlocal interactions due to fluctuating dipoles reduces the charge correlations at the interface of H-intercalated epitaxial graphene systems.  When considering the $(6\sqrt3\times6\sqrt3)R30^{\circ}$ model with the 338 graphene atoms in the supercell, the sublattice differences get almost averaged out and all functionals converge in a minimal Dirac-point perturbation, which is consistent with the experiment. It should be noted here that this effect is inherent with nonuniform electrostatic potentials induced by substrates with atomic arrangements that are unequal to that of graphene\cite{2008PhRvL.100q6802K,2011PhRvL.106j6801N}. However, its quantitative manifestation in the H-passivated SiC(0001) case, as obtained by the $(6\sqrt3\times6\sqrt3)R30^{\circ}$ cell, is negligible. A second issue worth mentioning for the $(\sqrt3\times\sqrt3)R30^{\circ}$ case is the shift of the graphene eigenstates with respect to the substrate levels due to fictitious strain, which results in a positioning of the Dirac cone closer to the valence band of SiC with respect to the bigger supercell. Nonetheless, all models converge in the designation of charge neutrality at the Dirac point. This aspect can be twofold commented: on one hand the absence of an effective doping in this system shows that the H-passivated SiC(0001) surface looses the characteristics of the free polar SiC(0001) surface and does not introduce Fermi-level-pinning midgap states in the electronic structure. On the other hand, the simulation of an extremely  thin SiC slab neglects the formation of intrinsic dipoles formed at the (0001) direction of hexagonal SiC polytypes and may undermine their doping contributions. In this sense, the results presented in this paragraph should be more suitable for cubic, rather than hexagonal SiC. 

\section{Lithium intercalation}
\label{lithium}

\begin{figure}
	\centering
		\includegraphics[width=\columnwidth]{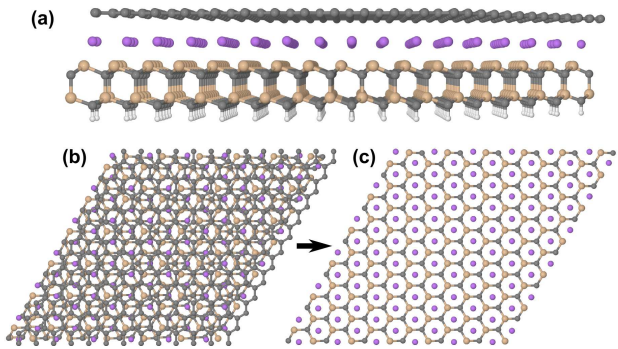}
	\caption{(Color online) (a) Side view of the relaxed Li-intercalated epitaxial graphene system. (b) Top view of the Li-intercalated system. (c) Top view showing only the Li atoms that occupy the hollow positions of the first SiC bilayer.}
	\label{fig:lithium}
\end{figure}

The main problem with the use of atomic hydrogen in the chemical detachment of the graphene/SiC interface is that it can also functionalize the graphene layer itself \cite{2009JChPh.130e4704C}, giving rise to $sp^3$-type defects. Such defects are at the origin of transport pseudogaps arround the Dirac point during electrical conduction and can substantially increase the electronic localization and, hence, the electrical disorder\cite{2010PhRvB..81s3412B}. A possible solution towards this direction could be the intercalation with elements that do not disturb the planarity and the $sp^2$ character of the graphene sheet. Lithium fulfills this prerequisite \cite{2004PhRvB..70l5422K}. Experimentally Li-intercalation has been proposed by Virojanadara \textit{et al.} \cite{2010PhRvB..82t5402V} who showed that Li, as a highly reactive lightweight monovalent metal, can penetrate the strongly-bound carbon-rich layer and position on top of the SiC substrate. Motivated by such experiment, we model Li-intercalated epitaxial graphene systems considering an equivalent to H complete coverage of the SiC surface. Structurally, also in this case the carbon interface layer relaxes into a flat graphene position with a vertical distance of 2.37 \AA{} from the Li adlayer and 4.41 \AA{} from the Si atoms of the substrate [Fig. \ref{fig:lithium}(a)]. However, as a first difference with respect to the H case, we find that the minimum-energy configuration for the relaxation of Li atoms is at the hollow positions of the first substrate bilayer [Fig. \ref{fig:lithium}(b-c)]. The reason of this symmetry originates from the bonding interactions between the Li adlayer and the substrate. We find an important charge transfer from the Li atoms to a region between Li and the Si atoms of the substrate's surface, localized on the top of the Si atomic positions. Such configuration guarantees the pure $sp^3$ hybridization of the substrate, while it leaves a  cation at the location of the Li atom. The unequal chemisorption of H and Li on the graphene/SiC interface could be also reflected in a different sub-graphene inter-diffusion mechanism during the intercalation process.


\begin{figure*}
	\centering
		\includegraphics[width=2\columnwidth]{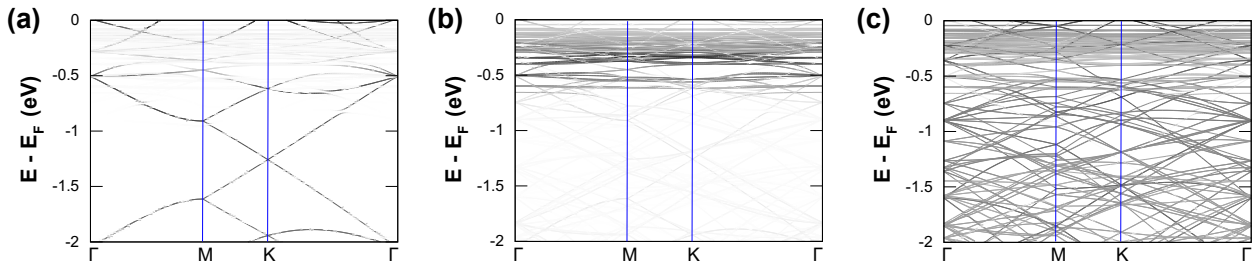}
	\caption{(Color online) Orbital-resolved band structure for the  Li-intercalated epitaxial graphene system, showing the contributions of the (a) $2p_z$ orbitals originating from the graphene C atoms, (b) all orbitals originating from Li atoms and (c) all orbitals originating from the substrate C/Si atoms. The z-axis is perpendicular to the graphene plain.}
	\label{fig:Li-bands}
\end{figure*}


The electronic band structure obtained by this composite system shows highly-concentrated bands throughout the hole region of the spectrum that are difficult to interpret. In this sense an orbital-resolved analysis is necessary for the determination of the local contributions in the total electronic structure. Considering therefore the contribution of the $2p_z$ orbitals that originate from the graphene layer, we obtain a band structure that is equal to that of ideal graphene for a large energy range around the Dirac point [Fig. \ref{fig:Li-bands}(a)]. This result denotes that the detachment from the substrate also in the case of Li intercalation is almost complete, while unlike the H-intercalated case we do not obtain any interference at the Dirac point due to a better screening of the substrate from the Li ionic cores. The important aspect though is that the graphene sheet is highly electron-doped (as also calculated for a low-coverage Au intercalation \cite{2011Nanot..22A5704C}), with the Dirac cone lying $\sim 1.3$ eV below the Fermi level of the heterosystem. This high effective doping is due to the electrostatic interactions between the graphene layer and the Li cations, which gives rise to a shift of the graphene bands within the valence band region of the SiC substrate [Fig. \ref{fig:Li-bands}(c)], making the whole system electrically active at the vicinity of the Dirac point. The analysis of the electronic Hamiltonian of the composite system here shows an important decrease of the on-site energies of the C atoms with respect to the case of H-intercalation. This aspect is important if we consider that the actual intercalation process may not be uniform throughout the entire breadth of the graphene layer. In this case we can expect an Anderson-type disorder arising from the energetic mismatch between intercalated and non-intercalated regions. A further proof of the differences between the H and Li-intercalated systems results from the calculation of the binding energies of the graphene layer on the passivated substrate: here the LDA values are 0.06 eV and 0.108 eV per C atom respectively. Finally, it is interesting to note that the quasi-flat bands originating from the Li orbitals [Fig. \ref{fig:Li-bands}(b)] maintain a higher-energy position with respect to the Dirac cone and, hence, should not interfere during electrical conduction around Dirac-point energies.  

\section{Discussion}
\label{discussion}

The extrapolation of device-requested characteristics, like high electron mobilities and densities, from epitaxial graphene systems grown on the Si face of SiC makes necessary the passivation of the interface between graphene and SiC. Towards this objective, intercalation techniques with various functional elements have been proposed in the literature. In this article we have studied within a DFT framework the electronic structure properties of the non-intercalated as well as intercalated interfaces between graphene and the Si face of SiC, starting from the actual $(6\sqrt3\times6\sqrt3)R30^{\circ}$ reconstruction of the graphene/SiC heterosystem. We identified some issues related to the graphene/SiC coupling, like the presence of midgap interface states within the SiC bandgap, which could give rise to plausible faults in the expected electrical behavior. We thereon studied two different intercalated systems with H and Li adatoms respectively, and saw that although in both cases the strongly-bound carbon-rich layer gets detached from the substrate, different electronic properties distinguish the two structures: in the case of H intercalation we observed a minimal perturbation at the Dirac point as a consequence of substrate-induced interference from localized Si-H dipoles, while in the case of Li intercalation we obtained a highly $n$-doped ideal graphene band structure with the position of the Dirac cone within the valence band of the SiC substrate. A critical comparison between these two examples unfolds the potentiality of substrate engineering in epitaxial graphene on SiC. 

It can be argued that research on intercalation processes for the use of epitaxial graphene in applications should focus on: (a) a full and uniform substrate coverage, (b)  an absence of functionalized areas on the graphene layer, and (c) an absence of non-graphene states around the Dirac point. Moreover, the level of doping can be appropriately used in order to engineer the active regions for device components. Within this context, both covalent and metallic intercalation mechanisms could be useful for future epitaxial graphene device integration.

\begin{acknowledgments}
This work was partially supported by the European Science Foundation (ESF) under the EUROCORES Program EuroGRAPHENE CRP GRAPHIC-RF. Computations have been performed at the CINECA supercomputing facilities under project TRAGRAPH.
\end{acknowledgments}

\bibliography{intercalated-long}

\begin{thebibliography}{36}%
\makeatletter
\providecommand \@ifxundefined [1]{%
 \@ifx{#1\undefined}
}%
\providecommand \@ifnum [1]{%
 \ifnum #1\expandafter \@firstoftwo
 \else \expandafter \@secondoftwo
 \fi
}%
\providecommand \@ifx [1]{%
 \ifx #1\expandafter \@firstoftwo
 \else \expandafter \@secondoftwo
 \fi
}%
\providecommand \natexlab [1]{#1}%
\providecommand \enquote  [1]{``#1''}%
\providecommand \bibnamefont  [1]{#1}%
\providecommand \bibfnamefont [1]{#1}%
\providecommand \citenamefont [1]{#1}%
\providecommand \href@noop [0]{\@secondoftwo}%
\providecommand \href [0]{\begingroup \@sanitize@url \@href}%
\providecommand \@href[1]{\@@startlink{#1}\@@href}%
\providecommand \@@href[1]{\endgroup#1\@@endlink}%
\providecommand \@sanitize@url [0]{\catcode `\\12\catcode `\$12\catcode
  `\&12\catcode `\#12\catcode `\^12\catcode `\_12\catcode `\%12\relax}%
\providecommand \@@startlink[1]{}%
\providecommand \@@endlink[0]{}%
\providecommand \url  [0]{\begingroup\@sanitize@url \@url }%
\providecommand \@url [1]{\endgroup\@href {#1}{\urlprefix }}%
\providecommand \urlprefix  [0]{URL }%
\providecommand \Eprint [0]{\href }%
\providecommand \doibase [0]{http://dx.doi.org/}%
\providecommand \selectlanguage [0]{\@gobble}%
\providecommand \bibinfo  [0]{\@secondoftwo}%
\providecommand \bibfield  [0]{\@secondoftwo}%
\providecommand \translation [1]{[#1]}%
\providecommand \BibitemOpen [0]{}%
\providecommand \bibitemStop [0]{}%
\providecommand \bibitemNoStop [0]{.\EOS\space}%
\providecommand \EOS [0]{\spacefactor3000\relax}%
\providecommand \BibitemShut  [1]{\csname bibitem#1\endcsname}%
\let\auto@bib@innerbib\@empty
\bibitem [{\citenamefont {{Castro Neto}}\ \emph {et~al.}(2009)\citenamefont
  {{Castro Neto}}, \citenamefont {{Guinea}}, \citenamefont {{Peres}},
  \citenamefont {{Novoselov}},\ and\ \citenamefont
  {{Geim}}}]{2009RvMP...81..109C}%
  \BibitemOpen
  \bibfield  {author} {\bibinfo {author} {\bibfnamefont {A.~H.}\ \bibnamefont
  {{Castro Neto}}}, \bibinfo {author} {\bibfnamefont {F.}~\bibnamefont
  {{Guinea}}}, \bibinfo {author} {\bibfnamefont {N.~M.~R.}\ \bibnamefont
  {{Peres}}}, \bibinfo {author} {\bibfnamefont {K.~S.}\ \bibnamefont
  {{Novoselov}}}, \ and\ \bibinfo {author} {\bibfnamefont {A.~K.}\ \bibnamefont
  {{Geim}}},\ }\href {\doibase 10.1103/RevModPhys.81.109} {\bibfield  {journal}
  {\bibinfo  {journal} {\rmp}\ }\textbf {\bibinfo {volume} {81}},\ \bibinfo
  {pages} {109} (\bibinfo {year} {2009})}\BibitemShut {NoStop}%
\bibitem [{\citenamefont {{Emtsev}}\ \emph {et~al.}(2009)\citenamefont
  {{Emtsev}}, \citenamefont {{Bostwick}}, \citenamefont {{Horn}}, \citenamefont
  {{Jobst}}, \citenamefont {{Kellogg}}, \citenamefont {{Ley}}, \citenamefont
  {{McChesney}}, \citenamefont {{Ohta}}, \citenamefont {{Reshanov}},
  \citenamefont {{R{\"o}hrl}}, \citenamefont {{Rotenberg}}, \citenamefont
  {{Schmid}}, \citenamefont {{Waldmann}}, \citenamefont {{Weber}},\ and\
  \citenamefont {{Seyller}}}]{2009NatMa...8..203E}%
  \BibitemOpen
  \bibfield  {author} {\bibinfo {author} {\bibfnamefont {K.~V.}\ \bibnamefont
  {{Emtsev}}}, \bibinfo {author} {\bibfnamefont {A.}~\bibnamefont
  {{Bostwick}}}, \bibinfo {author} {\bibfnamefont {K.}~\bibnamefont {{Horn}}},
  \bibinfo {author} {\bibfnamefont {J.}~\bibnamefont {{Jobst}}}, \bibinfo
  {author} {\bibfnamefont {G.~L.}\ \bibnamefont {{Kellogg}}}, \bibinfo {author}
  {\bibfnamefont {L.}~\bibnamefont {{Ley}}}, \bibinfo {author} {\bibfnamefont
  {J.~L.}\ \bibnamefont {{McChesney}}}, \bibinfo {author} {\bibfnamefont
  {T.}~\bibnamefont {{Ohta}}}, \bibinfo {author} {\bibfnamefont {S.~A.}\
  \bibnamefont {{Reshanov}}}, \bibinfo {author} {\bibfnamefont
  {J.}~\bibnamefont {{R{\"o}hrl}}}, \bibinfo {author} {\bibfnamefont
  {E.}~\bibnamefont {{Rotenberg}}}, \bibinfo {author} {\bibfnamefont {A.~K.}\
  \bibnamefont {{Schmid}}}, \bibinfo {author} {\bibfnamefont {D.}~\bibnamefont
  {{Waldmann}}}, \bibinfo {author} {\bibfnamefont {H.~B.}\ \bibnamefont
  {{Weber}}}, \ and\ \bibinfo {author} {\bibfnamefont {T.}~\bibnamefont
  {{Seyller}}},\ }\href {\doibase 10.1038/nmat2382} {\bibfield  {journal}
  {\bibinfo  {journal} {Nature Materials}\ }\textbf {\bibinfo {volume} {8}},\
  \bibinfo {pages} {203} (\bibinfo {year} {2009})}\BibitemShut {NoStop}%
\bibitem [{\citenamefont {{Dimitrakopoulos}}\ \emph {et~al.}(2010)\citenamefont
  {{Dimitrakopoulos}}, \citenamefont {{Lin}}, \citenamefont {{Grill}},
  \citenamefont {{Farmer}}, \citenamefont {{Freitag}}, \citenamefont {{Sun}},
  \citenamefont {{Han}}, \citenamefont {{Chen}}, \citenamefont {{Jenkins}},
  \citenamefont {{Zhu}}, \citenamefont {{Liu}}, \citenamefont {{McArdle}},
  \citenamefont {{Ott}}, \citenamefont {{Wisnieff}},\ and\ \citenamefont
  {{Avouris}}}]{2010JVSTB..28..985D}%
  \BibitemOpen
  \bibfield  {author} {\bibinfo {author} {\bibfnamefont {C.}~\bibnamefont
  {{Dimitrakopoulos}}}, \bibinfo {author} {\bibfnamefont {Y.-M.}\ \bibnamefont
  {{Lin}}}, \bibinfo {author} {\bibfnamefont {A.}~\bibnamefont {{Grill}}},
  \bibinfo {author} {\bibfnamefont {D.~B.}\ \bibnamefont {{Farmer}}}, \bibinfo
  {author} {\bibfnamefont {M.}~\bibnamefont {{Freitag}}}, \bibinfo {author}
  {\bibfnamefont {Y.}~\bibnamefont {{Sun}}}, \bibinfo {author} {\bibfnamefont
  {S.-J.}\ \bibnamefont {{Han}}}, \bibinfo {author} {\bibfnamefont
  {Z.}~\bibnamefont {{Chen}}}, \bibinfo {author} {\bibfnamefont {K.~A.}\
  \bibnamefont {{Jenkins}}}, \bibinfo {author} {\bibfnamefont {Y.}~\bibnamefont
  {{Zhu}}}, \bibinfo {author} {\bibfnamefont {Z.}~\bibnamefont {{Liu}}},
  \bibinfo {author} {\bibfnamefont {T.~J.}\ \bibnamefont {{McArdle}}}, \bibinfo
  {author} {\bibfnamefont {J.~A.}\ \bibnamefont {{Ott}}}, \bibinfo {author}
  {\bibfnamefont {R.}~\bibnamefont {{Wisnieff}}}, \ and\ \bibinfo {author}
  {\bibfnamefont {P.}~\bibnamefont {{Avouris}}},\ }\href {\doibase
  10.1116/1.3480961} {\bibfield  {journal} {\bibinfo  {journal} {Journal of
  Vacuum Science Technology B}\ }\textbf {\bibinfo {volume} {28}},\ \bibinfo
  {pages} {985} (\bibinfo {year} {2010})}\BibitemShut {NoStop}%
\bibitem [{\citenamefont {{Jernigan}}\ \emph {et~al.}(2009)\citenamefont
  {{Jernigan}}, \citenamefont {{Vanmil}}, \citenamefont {{Tedesco}},
  \citenamefont {{Tischler}}, \citenamefont {{Glaser}}, \citenamefont
  {{Davidson}}, \citenamefont {{Campbell}},\ and\ \citenamefont
  {{Gaskill}}}]{2009NanoL...9.2605J}%
  \BibitemOpen
  \bibfield  {author} {\bibinfo {author} {\bibfnamefont {G.~G.}\ \bibnamefont
  {{Jernigan}}}, \bibinfo {author} {\bibfnamefont {B.~L.}\ \bibnamefont
  {{Vanmil}}}, \bibinfo {author} {\bibfnamefont {J.~L.}\ \bibnamefont
  {{Tedesco}}}, \bibinfo {author} {\bibfnamefont {J.~G.}\ \bibnamefont
  {{Tischler}}}, \bibinfo {author} {\bibfnamefont {E.~R.}\ \bibnamefont
  {{Glaser}}}, \bibinfo {author} {\bibfnamefont {A.}~\bibnamefont {{Davidson}},
  \bibfnamefont {III}}, \bibinfo {author} {\bibfnamefont {P.~M.}\ \bibnamefont
  {{Campbell}}}, \ and\ \bibinfo {author} {\bibfnamefont {D.~K.}\ \bibnamefont
  {{Gaskill}}},\ }\href {\doibase 10.1021/nl900803z} {\bibfield  {journal}
  {\bibinfo  {journal} {Nano Letters}\ }\textbf {\bibinfo {volume} {9}},\
  \bibinfo {pages} {2605} (\bibinfo {year} {2009})}\BibitemShut {NoStop}%
\bibitem [{\citenamefont {{Hass}}\ \emph {et~al.}(2008)\citenamefont {{Hass}},
  \citenamefont {{de Heer}},\ and\ \citenamefont
  {{Conrad}}}]{2008JPCM...20F3202H}%
  \BibitemOpen
  \bibfield  {author} {\bibinfo {author} {\bibfnamefont {J.}~\bibnamefont
  {{Hass}}}, \bibinfo {author} {\bibfnamefont {W.~A.}\ \bibnamefont {{de
  Heer}}}, \ and\ \bibinfo {author} {\bibfnamefont {E.~H.}\ \bibnamefont
  {{Conrad}}},\ }\href {\doibase 10.1088/0953-8984/20/32/323202} {\bibfield
  {journal} {\bibinfo  {journal} {Journal of Physics Condensed Matter}\
  }\textbf {\bibinfo {volume} {20}},\ \bibinfo {pages} {323202} (\bibinfo
  {year} {2008})}\BibitemShut {NoStop}%
\bibitem [{\citenamefont {{Ouerghi}}\ \emph {et~al.}(2010)\citenamefont
  {{Ouerghi}}, \citenamefont {{Kahouli}}, \citenamefont {{Lucot}},
  \citenamefont {{Portail}}, \citenamefont {{Travers}}, \citenamefont
  {{Gierak}}, \citenamefont {{Penuelas}}, \citenamefont {{Jegou}},
  \citenamefont {{Shukla}}, \citenamefont {{Chassagne}},\ and\ \citenamefont
  {{Zielinski}}}]{2010ApPhL..96s1910O}%
  \BibitemOpen
  \bibfield  {author} {\bibinfo {author} {\bibfnamefont {A.}~\bibnamefont
  {{Ouerghi}}}, \bibinfo {author} {\bibfnamefont {A.}~\bibnamefont
  {{Kahouli}}}, \bibinfo {author} {\bibfnamefont {D.}~\bibnamefont {{Lucot}}},
  \bibinfo {author} {\bibfnamefont {M.}~\bibnamefont {{Portail}}}, \bibinfo
  {author} {\bibfnamefont {L.}~\bibnamefont {{Travers}}}, \bibinfo {author}
  {\bibfnamefont {J.}~\bibnamefont {{Gierak}}}, \bibinfo {author}
  {\bibfnamefont {J.}~\bibnamefont {{Penuelas}}}, \bibinfo {author}
  {\bibfnamefont {P.}~\bibnamefont {{Jegou}}}, \bibinfo {author} {\bibfnamefont
  {A.}~\bibnamefont {{Shukla}}}, \bibinfo {author} {\bibfnamefont
  {T.}~\bibnamefont {{Chassagne}}}, \ and\ \bibinfo {author} {\bibfnamefont
  {M.}~\bibnamefont {{Zielinski}}},\ }\href {\doibase 10.1063/1.3427406}
  {\bibfield  {journal} {\bibinfo  {journal} {\apl}\ }\textbf {\bibinfo
  {volume} {96}},\ \bibinfo {pages} {191910} (\bibinfo {year}
  {2010})}\BibitemShut {NoStop}%
\bibitem [{\citenamefont {{Pankratov}}\ \emph {et~al.}(2010)\citenamefont
  {{Pankratov}}, \citenamefont {{Hensel}},\ and\ \citenamefont
  {{Bockstedte}}}]{2010PhRvB..82l1416P}%
  \BibitemOpen
  \bibfield  {author} {\bibinfo {author} {\bibfnamefont {O.}~\bibnamefont
  {{Pankratov}}}, \bibinfo {author} {\bibfnamefont {S.}~\bibnamefont
  {{Hensel}}}, \ and\ \bibinfo {author} {\bibfnamefont {M.}~\bibnamefont
  {{Bockstedte}}},\ }\href {\doibase 10.1103/PhysRevB.82.121416} {\bibfield
  {journal} {\bibinfo  {journal} {\prb}\ }\textbf {\bibinfo {volume} {82}},\
  \bibinfo {pages} {121416} (\bibinfo {year} {2010})}\BibitemShut {NoStop}%
\bibitem [{\citenamefont {{Deretzis}}\ and\ \citenamefont {{La
  Magna}}(2011)}]{2011ApPhL..98b3113D}%
  \BibitemOpen
  \bibfield  {author} {\bibinfo {author} {\bibfnamefont {I.}~\bibnamefont
  {{Deretzis}}}\ and\ \bibinfo {author} {\bibfnamefont {A.}~\bibnamefont {{La
  Magna}}},\ }\href {\doibase 10.1063/1.3543847} {\bibfield  {journal}
  {\bibinfo  {journal} {\apl}\ }\textbf {\bibinfo {volume} {98}},\ \bibinfo
  {pages} {023113} (\bibinfo {year} {2011})}\BibitemShut {NoStop}%
\bibitem [{\citenamefont {{Wu}}\ \emph {et~al.}(2009)\citenamefont {{Wu}},
  \citenamefont {{Hu}}, \citenamefont {{Ruan}}, \citenamefont {{Madiomanana}},
  \citenamefont {{Hankinson}}, \citenamefont {{Sprinkle}}, \citenamefont
  {{Berger}},\ and\ \citenamefont {{de Heer}}}]{2009ApPhL..95v3108W}%
  \BibitemOpen
  \bibfield  {author} {\bibinfo {author} {\bibfnamefont {X.}~\bibnamefont
  {{Wu}}}, \bibinfo {author} {\bibfnamefont {Y.}~\bibnamefont {{Hu}}}, \bibinfo
  {author} {\bibfnamefont {M.}~\bibnamefont {{Ruan}}}, \bibinfo {author}
  {\bibfnamefont {N.~K.}\ \bibnamefont {{Madiomanana}}}, \bibinfo {author}
  {\bibfnamefont {J.}~\bibnamefont {{Hankinson}}}, \bibinfo {author}
  {\bibfnamefont {M.}~\bibnamefont {{Sprinkle}}}, \bibinfo {author}
  {\bibfnamefont {C.}~\bibnamefont {{Berger}}}, \ and\ \bibinfo {author}
  {\bibfnamefont {W.~A.}\ \bibnamefont {{de Heer}}},\ }\href {\doibase
  10.1063/1.3266524} {\bibfield  {journal} {\bibinfo  {journal} {\apl}\
  }\textbf {\bibinfo {volume} {95}},\ \bibinfo {pages} {223108} (\bibinfo
  {year} {2009})}\BibitemShut {NoStop}%
\bibitem [{\citenamefont {{Kim}}\ \emph {et~al.}(2008)\citenamefont {{Kim}},
  \citenamefont {{Ihm}}, \citenamefont {{Choi}},\ and\ \citenamefont
  {{Son}}}]{2008PhRvL.100q6802K}%
  \BibitemOpen
  \bibfield  {author} {\bibinfo {author} {\bibfnamefont {S.}~\bibnamefont
  {{Kim}}}, \bibinfo {author} {\bibfnamefont {J.}~\bibnamefont {{Ihm}}},
  \bibinfo {author} {\bibfnamefont {H.~J.}\ \bibnamefont {{Choi}}}, \ and\
  \bibinfo {author} {\bibfnamefont {Y.-W.}\ \bibnamefont {{Son}}},\ }\href
  {\doibase 10.1103/PhysRevLett.100.176802} {\bibfield  {journal} {\bibinfo
  {journal} {\prl}\ }\textbf {\bibinfo {volume} {100}},\ \bibinfo {pages}
  {176802} (\bibinfo {year} {2008})}\BibitemShut {NoStop}%
\bibitem [{\citenamefont {{Sonde}}\ \emph {et~al.}(2010)\citenamefont
  {{Sonde}}, \citenamefont {{Giannazzo}}, \citenamefont {{Vecchio}},
  \citenamefont {{Yakimova}}, \citenamefont {{Rimini}},\ and\ \citenamefont
  {{Raineri}}}]{2010ApPhL..97m2101S}%
  \BibitemOpen
  \bibfield  {author} {\bibinfo {author} {\bibfnamefont {S.}~\bibnamefont
  {{Sonde}}}, \bibinfo {author} {\bibfnamefont {F.}~\bibnamefont
  {{Giannazzo}}}, \bibinfo {author} {\bibfnamefont {C.}~\bibnamefont
  {{Vecchio}}}, \bibinfo {author} {\bibfnamefont {R.}~\bibnamefont
  {{Yakimova}}}, \bibinfo {author} {\bibfnamefont {E.}~\bibnamefont
  {{Rimini}}}, \ and\ \bibinfo {author} {\bibfnamefont {V.}~\bibnamefont
  {{Raineri}}},\ }\href {\doibase 10.1063/1.3489942} {\bibfield  {journal}
  {\bibinfo  {journal} {\apl}\ }\textbf {\bibinfo {volume} {97}},\ \bibinfo
  {pages} {132101} (\bibinfo {year} {2010})}\BibitemShut {NoStop}%
\bibitem [{\citenamefont {{Speck}}\ \emph {et~al.}(2011)\citenamefont
  {{Speck}}, \citenamefont {{Jobst}}, \citenamefont {{Fromm}}, \citenamefont
  {{Ostler}}, \citenamefont {{Waldmann}}, \citenamefont {{Hundhausen}},
  \citenamefont {{Weber}},\ and\ \citenamefont
  {{Seyller}}}]{2011arXiv1103.3997S}%
  \BibitemOpen
  \bibfield  {author} {\bibinfo {author} {\bibfnamefont {F.}~\bibnamefont
  {{Speck}}}, \bibinfo {author} {\bibfnamefont {J.}~\bibnamefont {{Jobst}}},
  \bibinfo {author} {\bibfnamefont {F.}~\bibnamefont {{Fromm}}}, \bibinfo
  {author} {\bibfnamefont {M.}~\bibnamefont {{Ostler}}}, \bibinfo {author}
  {\bibfnamefont {D.}~\bibnamefont {{Waldmann}}}, \bibinfo {author}
  {\bibfnamefont {M.}~\bibnamefont {{Hundhausen}}}, \bibinfo {author}
  {\bibfnamefont {H.~B.}\ \bibnamefont {{Weber}}}, \ and\ \bibinfo {author}
  {\bibfnamefont {T.}~\bibnamefont {{Seyller}}},\ }\href {\doibase
  10.1063/1.3643034} {\bibfield  {journal} {\bibinfo  {journal} {\apl}\
  }\textbf {\bibinfo {volume} {99}},\ \bibinfo {eid} {122106} (\bibinfo {year}
  {2011})}\BibitemShut {NoStop}%
\bibitem [{\citenamefont {{Riedl}}\ \emph {et~al.}(2009)\citenamefont
  {{Riedl}}, \citenamefont {{Coletti}}, \citenamefont {{Iwasaki}},
  \citenamefont {{Zakharov}},\ and\ \citenamefont
  {{Starke}}}]{2009PhRvL.103x6804R}%
  \BibitemOpen
  \bibfield  {author} {\bibinfo {author} {\bibfnamefont {C.}~\bibnamefont
  {{Riedl}}}, \bibinfo {author} {\bibfnamefont {C.}~\bibnamefont {{Coletti}}},
  \bibinfo {author} {\bibfnamefont {T.}~\bibnamefont {{Iwasaki}}}, \bibinfo
  {author} {\bibfnamefont {A.~A.}\ \bibnamefont {{Zakharov}}}, \ and\ \bibinfo
  {author} {\bibfnamefont {U.}~\bibnamefont {{Starke}}},\ }\href {\doibase
  10.1103/PhysRevLett.103.246804} {\bibfield  {journal} {\bibinfo  {journal}
  {\prl}\ }\textbf {\bibinfo {volume} {103}},\ \bibinfo {pages} {246804}
  (\bibinfo {year} {2009})}\BibitemShut {NoStop}%
\bibitem [{\citenamefont {{Virojanadara}}\ \emph
  {et~al.}(2010{\natexlab{a}})\citenamefont {{Virojanadara}}, \citenamefont
  {{Watcharinyanon}}, \citenamefont {{Zakharov}},\ and\ \citenamefont
  {{Johansson}}}]{2010PhRvB..82t5402V}%
  \BibitemOpen
  \bibfield  {author} {\bibinfo {author} {\bibfnamefont {C.}~\bibnamefont
  {{Virojanadara}}}, \bibinfo {author} {\bibfnamefont {S.}~\bibnamefont
  {{Watcharinyanon}}}, \bibinfo {author} {\bibfnamefont {A.~A.}\ \bibnamefont
  {{Zakharov}}}, \ and\ \bibinfo {author} {\bibfnamefont {L.~I.}\ \bibnamefont
  {{Johansson}}},\ }\href {\doibase 10.1103/PhysRevB.82.205402} {\bibfield
  {journal} {\bibinfo  {journal} {\prb}\ }\textbf {\bibinfo {volume} {82}},\
  \bibinfo {pages} {205402} (\bibinfo {year} {2010}{\natexlab{a}})}\BibitemShut
  {NoStop}%
\bibitem [{\citenamefont {{Walter}}\ \emph {et~al.}(2011)\citenamefont
  {{Walter}}, \citenamefont {{Jeon}}, \citenamefont {{Bostwick}}, \citenamefont
  {{Speck}}, \citenamefont {{Ostler}}, \citenamefont {{Seyller}}, \citenamefont
  {{Moreschini}}, \citenamefont {{Kim}}, \citenamefont {{Chang}}, \citenamefont
  {{Horn}},\ and\ \citenamefont {{Rotenberg}}}]{2011ApPhL..98r4102W}%
  \BibitemOpen
  \bibfield  {author} {\bibinfo {author} {\bibfnamefont {A.~L.}\ \bibnamefont
  {{Walter}}}, \bibinfo {author} {\bibfnamefont {K.-J.}\ \bibnamefont
  {{Jeon}}}, \bibinfo {author} {\bibfnamefont {A.}~\bibnamefont {{Bostwick}}},
  \bibinfo {author} {\bibfnamefont {F.}~\bibnamefont {{Speck}}}, \bibinfo
  {author} {\bibfnamefont {M.}~\bibnamefont {{Ostler}}}, \bibinfo {author}
  {\bibfnamefont {T.}~\bibnamefont {{Seyller}}}, \bibinfo {author}
  {\bibfnamefont {L.}~\bibnamefont {{Moreschini}}}, \bibinfo {author}
  {\bibfnamefont {Y.~S.}\ \bibnamefont {{Kim}}}, \bibinfo {author}
  {\bibfnamefont {Y.~J.}\ \bibnamefont {{Chang}}}, \bibinfo {author}
  {\bibfnamefont {K.}~\bibnamefont {{Horn}}}, \ and\ \bibinfo {author}
  {\bibfnamefont {E.}~\bibnamefont {{Rotenberg}}},\ }\href {\doibase
  10.1063/1.3586256} {\bibfield  {journal} {\bibinfo  {journal} {\apl}\
  }\textbf {\bibinfo {volume} {98}},\ \bibinfo {pages} {184102} (\bibinfo
  {year} {2011})}\BibitemShut {NoStop}%
\bibitem [{\citenamefont {{Gierz}}\ \emph {et~al.}(2010)\citenamefont
  {{Gierz}}, \citenamefont {{Suzuki}}, \citenamefont {{Weitz}}, \citenamefont
  {{Lee}}, \citenamefont {{Krauss}}, \citenamefont {{Riedl}}, \citenamefont
  {{Starke}}, \citenamefont {{H{\"o}chst}}, \citenamefont {{Smet}},
  \citenamefont {{Ast}},\ and\ \citenamefont {{Kern}}}]{2010PhRvB..81w5408G}%
  \BibitemOpen
  \bibfield  {author} {\bibinfo {author} {\bibfnamefont {I.}~\bibnamefont
  {{Gierz}}}, \bibinfo {author} {\bibfnamefont {T.}~\bibnamefont {{Suzuki}}},
  \bibinfo {author} {\bibfnamefont {R.~T.}\ \bibnamefont {{Weitz}}}, \bibinfo
  {author} {\bibfnamefont {D.~S.}\ \bibnamefont {{Lee}}}, \bibinfo {author}
  {\bibfnamefont {B.}~\bibnamefont {{Krauss}}}, \bibinfo {author}
  {\bibfnamefont {C.}~\bibnamefont {{Riedl}}}, \bibinfo {author} {\bibfnamefont
  {U.}~\bibnamefont {{Starke}}}, \bibinfo {author} {\bibfnamefont
  {H.}~\bibnamefont {{H{\"o}chst}}}, \bibinfo {author} {\bibfnamefont {J.~H.}\
  \bibnamefont {{Smet}}}, \bibinfo {author} {\bibfnamefont {C.~R.}\
  \bibnamefont {{Ast}}}, \ and\ \bibinfo {author} {\bibfnamefont
  {K.}~\bibnamefont {{Kern}}},\ }\href {\doibase 10.1103/PhysRevB.81.235408}
  {\bibfield  {journal} {\bibinfo  {journal} {\prb}\ }\textbf {\bibinfo
  {volume} {81}},\ \bibinfo {pages} {235408} (\bibinfo {year}
  {2010})}\BibitemShut {NoStop}%
\bibitem [{\citenamefont {{Emtsev}}\ \emph {et~al.}(2011)\citenamefont
  {{Emtsev}}, \citenamefont {{Zakharov}}, \citenamefont {{Coletti}},
  \citenamefont {{Forti}},\ and\ \citenamefont
  {{Starke}}}]{2011PhRvB..84l5423E}%
  \BibitemOpen
  \bibfield  {author} {\bibinfo {author} {\bibfnamefont {K.~V.}\ \bibnamefont
  {{Emtsev}}}, \bibinfo {author} {\bibfnamefont {A.~A.}\ \bibnamefont
  {{Zakharov}}}, \bibinfo {author} {\bibfnamefont {C.}~\bibnamefont
  {{Coletti}}}, \bibinfo {author} {\bibfnamefont {S.}~\bibnamefont {{Forti}}},
  \ and\ \bibinfo {author} {\bibfnamefont {U.}~\bibnamefont {{Starke}}},\
  }\href {\doibase 10.1103/PhysRevB.84.125423} {\bibfield  {journal} {\bibinfo
  {journal} {\prb}\ }\textbf {\bibinfo {volume} {84}},\ \bibinfo {eid} {125423}
  (\bibinfo {year} {2011})}\BibitemShut {NoStop}%
\bibitem [{\citenamefont {{Soler}}\ \emph {et~al.}(2002)\citenamefont
  {{Soler}}, \citenamefont {{Artacho}}, \citenamefont {{Gale}}, \citenamefont
  {{Garc{\'{\i}}a}}, \citenamefont {{Junquera}}, \citenamefont
  {{Ordej{\'o}n}},\ and\ \citenamefont
  {{S{\'a}nchez-Portal}}}]{2002JPCM...14.2745S}%
  \BibitemOpen
  \bibfield  {author} {\bibinfo {author} {\bibfnamefont {J.~M.}\ \bibnamefont
  {{Soler}}}, \bibinfo {author} {\bibfnamefont {E.}~\bibnamefont {{Artacho}}},
  \bibinfo {author} {\bibfnamefont {J.~D.}\ \bibnamefont {{Gale}}}, \bibinfo
  {author} {\bibfnamefont {A.}~\bibnamefont {{Garc{\'{\i}}a}}}, \bibinfo
  {author} {\bibfnamefont {J.}~\bibnamefont {{Junquera}}}, \bibinfo {author}
  {\bibfnamefont {P.}~\bibnamefont {{Ordej{\'o}n}}}, \ and\ \bibinfo {author}
  {\bibfnamefont {D.}~\bibnamefont {{S{\'a}nchez-Portal}}},\ }\href {\doibase
  10.1088/0953-8984/14/11/302} {\bibfield  {journal} {\bibinfo  {journal}
  {Journal of Physics Condensed Matter}\ }\textbf {\bibinfo {volume} {14}},\
  \bibinfo {pages} {2745} (\bibinfo {year} {2002})}\BibitemShut {NoStop}%
\bibitem [{\citenamefont {{Perdew}}\ and\ \citenamefont
  {{Zunger}}(1981)}]{1981PhRvB..23.5048P}%
  \BibitemOpen
  \bibfield  {author} {\bibinfo {author} {\bibfnamefont {J.~P.}\ \bibnamefont
  {{Perdew}}}\ and\ \bibinfo {author} {\bibfnamefont {A.}~\bibnamefont
  {{Zunger}}},\ }\href {\doibase 10.1103/PhysRevB.23.5048} {\bibfield
  {journal} {\bibinfo  {journal} {\prb}\ }\textbf {\bibinfo {volume} {23}},\
  \bibinfo {pages} {5048} (\bibinfo {year} {1981})}\BibitemShut {NoStop}%
\bibitem [{\citenamefont {{Jayasekera}}\ \emph {et~al.}(2011)\citenamefont
  {{Jayasekera}}, \citenamefont {{Xu}}, \citenamefont {{Kim}},\ and\
  \citenamefont {{Nardelli}}}]{2011PhRvB..84c5442J}%
  \BibitemOpen
  \bibfield  {author} {\bibinfo {author} {\bibfnamefont {T.}~\bibnamefont
  {{Jayasekera}}}, \bibinfo {author} {\bibfnamefont {S.}~\bibnamefont {{Xu}}},
  \bibinfo {author} {\bibfnamefont {K.~W.}\ \bibnamefont {{Kim}}}, \ and\
  \bibinfo {author} {\bibfnamefont {M.~B.}\ \bibnamefont {{Nardelli}}},\ }\href
  {\doibase 10.1103/PhysRevB.84.035442} {\bibfield  {journal} {\bibinfo
  {journal} {\prb}\ }\textbf {\bibinfo {volume} {84}},\ \bibinfo {pages}
  {035442} (\bibinfo {year} {2011})}\BibitemShut {NoStop}%
\bibitem [{\citenamefont {{Perdew}}\ \emph {et~al.}(1996)\citenamefont
  {{Perdew}}, \citenamefont {{Burke}},\ and\ \citenamefont
  {{Ernzerhof}}}]{1996PhRvL..77.3865P}%
  \BibitemOpen
  \bibfield  {author} {\bibinfo {author} {\bibfnamefont {J.~P.}\ \bibnamefont
  {{Perdew}}}, \bibinfo {author} {\bibfnamefont {K.}~\bibnamefont {{Burke}}}, \
  and\ \bibinfo {author} {\bibfnamefont {M.}~\bibnamefont {{Ernzerhof}}},\
  }\href {\doibase 10.1103/PhysRevLett.77.3865} {\bibfield  {journal} {\bibinfo
   {journal} {\prl}\ }\textbf {\bibinfo {volume} {77}},\ \bibinfo {pages}
  {3865} (\bibinfo {year} {1996})}\BibitemShut {NoStop}%
\bibitem [{\citenamefont {{Dion}}\ \emph {et~al.}(2004)\citenamefont {{Dion}},
  \citenamefont {{Rydberg}}, \citenamefont {{Schr{\"o}der}}, \citenamefont
  {{Langreth}},\ and\ \citenamefont {{Lundqvist}}}]{2004PhRvL..92x6401D}%
  \BibitemOpen
  \bibfield  {author} {\bibinfo {author} {\bibfnamefont {M.}~\bibnamefont
  {{Dion}}}, \bibinfo {author} {\bibfnamefont {H.}~\bibnamefont {{Rydberg}}},
  \bibinfo {author} {\bibfnamefont {E.}~\bibnamefont {{Schr{\"o}der}}},
  \bibinfo {author} {\bibfnamefont {D.~C.}\ \bibnamefont {{Langreth}}}, \ and\
  \bibinfo {author} {\bibfnamefont {B.~I.}\ \bibnamefont {{Lundqvist}}},\
  }\href {\doibase 10.1103/PhysRevLett.92.246401} {\bibfield  {journal}
  {\bibinfo  {journal} {\prl}\ }\textbf {\bibinfo {volume} {92}},\ \bibinfo
  {pages} {246401} (\bibinfo {year} {2004})}\BibitemShut {NoStop}%
\bibitem [{\citenamefont {{Rom{\'a}n-P{\'e}rez}}\ and\ \citenamefont
  {{Soler}}(2009)}]{2009PhRvL.103i6102R}%
  \BibitemOpen
  \bibfield  {author} {\bibinfo {author} {\bibfnamefont {G.}~\bibnamefont
  {{Rom{\'a}n-P{\'e}rez}}}\ and\ \bibinfo {author} {\bibfnamefont {J.~M.}\
  \bibnamefont {{Soler}}},\ }\href {\doibase 10.1103/PhysRevLett.103.096102}
  {\bibfield  {journal} {\bibinfo  {journal} {\prl}\ }\textbf {\bibinfo
  {volume} {103}},\ \bibinfo {pages} {096102} (\bibinfo {year}
  {2009})}\BibitemShut {NoStop}%
\bibitem [{\citenamefont {{Deretzis}}\ and\ \citenamefont {{La
  Magna}}(2009)}]{2009ApPhL..95f3111D}%
  \BibitemOpen
  \bibfield  {author} {\bibinfo {author} {\bibfnamefont {I.}~\bibnamefont
  {{Deretzis}}}\ and\ \bibinfo {author} {\bibfnamefont {A.}~\bibnamefont {{La
  Magna}}},\ }\href {\doibase 10.1063/1.3202397} {\bibfield  {journal}
  {\bibinfo  {journal} {\apl}\ }\textbf {\bibinfo {volume} {95}},\ \bibinfo
  {pages} {063111} (\bibinfo {year} {2009})}\BibitemShut {NoStop}%
\bibitem [{\citenamefont {{Mattausch}}\ and\ \citenamefont
  {{Pankratov}}(2007)}]{2007PhRvL..99g6802M}%
  \BibitemOpen
  \bibfield  {author} {\bibinfo {author} {\bibfnamefont {A.}~\bibnamefont
  {{Mattausch}}}\ and\ \bibinfo {author} {\bibfnamefont {O.}~\bibnamefont
  {{Pankratov}}},\ }\href {\doibase 10.1103/PhysRevLett.99.076802} {\bibfield
  {journal} {\bibinfo  {journal} {\prl}\ }\textbf {\bibinfo {volume} {99}},\
  \bibinfo {pages} {076802} (\bibinfo {year} {2007})}\BibitemShut {NoStop}%
\bibitem [{\citenamefont {{Varchon}}\ \emph {et~al.}(2007)\citenamefont
  {{Varchon}}, \citenamefont {{Feng}}, \citenamefont {{Hass}}, \citenamefont
  {{Li}}, \citenamefont {{Nguyen}}, \citenamefont {{Naud}}, \citenamefont
  {{Mallet}}, \citenamefont {{Veuillen}}, \citenamefont {{Berger}},
  \citenamefont {{Conrad}},\ and\ \citenamefont
  {{Magaud}}}]{2007PhRvL..99l6805V}%
  \BibitemOpen
  \bibfield  {author} {\bibinfo {author} {\bibfnamefont {F.}~\bibnamefont
  {{Varchon}}}, \bibinfo {author} {\bibfnamefont {R.}~\bibnamefont {{Feng}}},
  \bibinfo {author} {\bibfnamefont {J.}~\bibnamefont {{Hass}}}, \bibinfo
  {author} {\bibfnamefont {X.}~\bibnamefont {{Li}}}, \bibinfo {author}
  {\bibfnamefont {B.~N.}\ \bibnamefont {{Nguyen}}}, \bibinfo {author}
  {\bibfnamefont {C.}~\bibnamefont {{Naud}}}, \bibinfo {author} {\bibfnamefont
  {P.}~\bibnamefont {{Mallet}}}, \bibinfo {author} {\bibfnamefont {J.-Y.}\
  \bibnamefont {{Veuillen}}}, \bibinfo {author} {\bibfnamefont
  {C.}~\bibnamefont {{Berger}}}, \bibinfo {author} {\bibfnamefont {E.~H.}\
  \bibnamefont {{Conrad}}}, \ and\ \bibinfo {author} {\bibfnamefont
  {L.}~\bibnamefont {{Magaud}}},\ }\href {\doibase
  10.1103/PhysRevLett.99.126805} {\bibfield  {journal} {\bibinfo  {journal}
  {\prl}\ }\textbf {\bibinfo {volume} {99}},\ \bibinfo {pages} {126805}
  (\bibinfo {year} {2007})}\BibitemShut {NoStop}%
\bibitem [{\citenamefont {{Troullier}}\ and\ \citenamefont
  {{Martins}}(1991)}]{1991PhRvB..43.1993T}%
  \BibitemOpen
  \bibfield  {author} {\bibinfo {author} {\bibfnamefont {N.}~\bibnamefont
  {{Troullier}}}\ and\ \bibinfo {author} {\bibfnamefont {J.~L.}\ \bibnamefont
  {{Martins}}},\ }\href {\doibase 10.1103/PhysRevB.43.1993} {\bibfield
  {journal} {\bibinfo  {journal} {\prb}\ }\textbf {\bibinfo {volume} {43}},\
  \bibinfo {pages} {1993} (\bibinfo {year} {1991})}\BibitemShut {NoStop}%
\bibitem [{\citenamefont {{Pollmann}}\ \emph {et~al.}(1997)\citenamefont
  {{Pollmann}}, \citenamefont {{Kr{\"u}ger}},\ and\ \citenamefont
  {{Sabisch}}}]{1997PSSBR.202..421P}%
  \BibitemOpen
  \bibfield  {author} {\bibinfo {author} {\bibfnamefont {J.}~\bibnamefont
  {{Pollmann}}}, \bibinfo {author} {\bibfnamefont {P.}~\bibnamefont
  {{Kr{\"u}ger}}}, \ and\ \bibinfo {author} {\bibfnamefont {M.}~\bibnamefont
  {{Sabisch}}},\ }\href {\doibase
  10.1002/1521-3951(199707)202:1<421::AID-PSSB421>3.0.CO;2-D} {\bibfield
  {journal} {\bibinfo  {journal} {Physica Status Solidi B Basic Research}\
  }\textbf {\bibinfo {volume} {202}},\ \bibinfo {pages} {421} (\bibinfo {year}
  {1997})}\BibitemShut {NoStop}%
\bibitem [{\citenamefont {{Tzalenchuk}}\ \emph {et~al.}(2010)\citenamefont
  {{Tzalenchuk}}, \citenamefont {{Lara-Avila}}, \citenamefont {{Kalaboukhov}},
  \citenamefont {{Paolillo}}, \citenamefont {{Syv{\"a}j{\"a}rvi}},
  \citenamefont {{Yakimova}}, \citenamefont {{Kazakova}}, \citenamefont
  {{Janssen}}, \citenamefont {{Fal'Ko}},\ and\ \citenamefont
  {{Kubatkin}}}]{2010NatNa...5..186T}%
  \BibitemOpen
  \bibfield  {author} {\bibinfo {author} {\bibfnamefont {A.}~\bibnamefont
  {{Tzalenchuk}}}, \bibinfo {author} {\bibfnamefont {S.}~\bibnamefont
  {{Lara-Avila}}}, \bibinfo {author} {\bibfnamefont {A.}~\bibnamefont
  {{Kalaboukhov}}}, \bibinfo {author} {\bibfnamefont {S.}~\bibnamefont
  {{Paolillo}}}, \bibinfo {author} {\bibfnamefont {M.}~\bibnamefont
  {{Syv{\"a}j{\"a}rvi}}}, \bibinfo {author} {\bibfnamefont {R.}~\bibnamefont
  {{Yakimova}}}, \bibinfo {author} {\bibfnamefont {O.}~\bibnamefont
  {{Kazakova}}}, \bibinfo {author} {\bibfnamefont {T.~J.~B.~M.}\ \bibnamefont
  {{Janssen}}}, \bibinfo {author} {\bibfnamefont {V.}~\bibnamefont {{Fal'Ko}}},
  \ and\ \bibinfo {author} {\bibfnamefont {S.}~\bibnamefont {{Kubatkin}}},\
  }\href {\doibase 10.1038/nnano.2009.474} {\bibfield  {journal} {\bibinfo
  {journal} {Nature Nanotechnology}\ }\textbf {\bibinfo {volume} {5}},\
  \bibinfo {pages} {186} (\bibinfo {year} {2010})}\BibitemShut {NoStop}%
\bibitem [{\citenamefont {{Virojanadara}}\ \emph
  {et~al.}(2010{\natexlab{b}})\citenamefont {{Virojanadara}}, \citenamefont
  {{Zakharov}}, \citenamefont {{Yakimova}},\ and\ \citenamefont
  {{Johansson}}}]{2010SurSc.604L...4V}%
  \BibitemOpen
  \bibfield  {author} {\bibinfo {author} {\bibfnamefont {C.}~\bibnamefont
  {{Virojanadara}}}, \bibinfo {author} {\bibfnamefont {A.~A.}\ \bibnamefont
  {{Zakharov}}}, \bibinfo {author} {\bibfnamefont {R.}~\bibnamefont
  {{Yakimova}}}, \ and\ \bibinfo {author} {\bibfnamefont {L.~I.}\ \bibnamefont
  {{Johansson}}},\ }\href {\doibase 10.1016/j.susc.2009.11.011} {\bibfield
  {journal} {\bibinfo  {journal} {Surface Science}\ }\textbf {\bibinfo {volume}
  {604}},\ \bibinfo {pages} {L4} (\bibinfo {year}
  {2010}{\natexlab{b}})}\BibitemShut {NoStop}%
\bibitem [{\citenamefont {{Cuong}}\ \emph {et~al.}(2011)\citenamefont
  {{Cuong}}, \citenamefont {{Otani}},\ and\ \citenamefont
  {{Okada}}}]{2011PhRvL.106j6801N}%
  \BibitemOpen
  \bibfield  {author} {\bibinfo {author} {\bibfnamefont {N.~T.}\ \bibnamefont
  {{Cuong}}}, \bibinfo {author} {\bibfnamefont {M.}~\bibnamefont {{Otani}}}, \
  and\ \bibinfo {author} {\bibfnamefont {S.}~\bibnamefont {{Okada}}},\ }\href
  {\doibase 10.1103/PhysRevLett.106.106801} {\bibfield  {journal} {\bibinfo
  {journal} {\prl}\ }\textbf {\bibinfo {volume} {106}},\ \bibinfo {pages}
  {106801} (\bibinfo {year} {2011})}\BibitemShut {NoStop}%
\bibitem [{\citenamefont {{Zhou}}\ \emph {et~al.}(2007)\citenamefont {{Zhou}},
  \citenamefont {{Gweon}}, \citenamefont {{Fedorov}}, \citenamefont {{First}},
  \citenamefont {{de Heer}}, \citenamefont {{Lee}}, \citenamefont {{Guinea}},
  \citenamefont {{Castro Neto}},\ and\ \citenamefont
  {{Lanzara}}}]{2007NatMa...6..770Z}%
  \BibitemOpen
  \bibfield  {author} {\bibinfo {author} {\bibfnamefont {S.~Y.}\ \bibnamefont
  {{Zhou}}}, \bibinfo {author} {\bibfnamefont {G.-H.}\ \bibnamefont {{Gweon}}},
  \bibinfo {author} {\bibfnamefont {A.~V.}\ \bibnamefont {{Fedorov}}}, \bibinfo
  {author} {\bibfnamefont {P.~N.}\ \bibnamefont {{First}}}, \bibinfo {author}
  {\bibfnamefont {W.~A.}\ \bibnamefont {{de Heer}}}, \bibinfo {author}
  {\bibfnamefont {D.-H.}\ \bibnamefont {{Lee}}}, \bibinfo {author}
  {\bibfnamefont {F.}~\bibnamefont {{Guinea}}}, \bibinfo {author}
  {\bibfnamefont {A.~H.}\ \bibnamefont {{Castro Neto}}}, \ and\ \bibinfo
  {author} {\bibfnamefont {A.}~\bibnamefont {{Lanzara}}},\ }\href {\doibase
  10.1038/nmat2003} {\bibfield  {journal} {\bibinfo  {journal} {Nature
  Materials}\ }\textbf {\bibinfo {volume} {6}},\ \bibinfo {pages} {770}
  (\bibinfo {year} {2007})}\BibitemShut {NoStop}%
\bibitem [{\citenamefont {{Casolo}}\ \emph {et~al.}(2009)\citenamefont
  {{Casolo}}, \citenamefont {{L{\o}vvik}}, \citenamefont {{Martinazzo}},\ and\
  \citenamefont {{Tantardini}}}]{2009JChPh.130e4704C}%
  \BibitemOpen
  \bibfield  {author} {\bibinfo {author} {\bibfnamefont {S.}~\bibnamefont
  {{Casolo}}}, \bibinfo {author} {\bibfnamefont {O.~M.}\ \bibnamefont
  {{L{\o}vvik}}}, \bibinfo {author} {\bibfnamefont {R.}~\bibnamefont
  {{Martinazzo}}}, \ and\ \bibinfo {author} {\bibfnamefont {G.~F.}\
  \bibnamefont {{Tantardini}}},\ }\href {\doibase 10.1063/1.3072333} {\bibfield
   {journal} {\bibinfo  {journal} {\jcp}\ }\textbf {\bibinfo {volume} {130}},\
  \bibinfo {pages} {054704} (\bibinfo {year} {2009})}\BibitemShut {NoStop}%
\bibitem [{\citenamefont {{Bang}}\ and\ \citenamefont
  {{Chang}}(2010)}]{2010PhRvB..81s3412B}%
  \BibitemOpen
  \bibfield  {author} {\bibinfo {author} {\bibfnamefont {J.}~\bibnamefont
  {{Bang}}}\ and\ \bibinfo {author} {\bibfnamefont {K.~J.}\ \bibnamefont
  {{Chang}}},\ }\href {\doibase 10.1103/PhysRevB.81.193412} {\bibfield
  {journal} {\bibinfo  {journal} {\prb}\ }\textbf {\bibinfo {volume} {81}},\
  \bibinfo {eid} {193412} (\bibinfo {year} {2010})}\BibitemShut {NoStop}%
\bibitem [{\citenamefont {{Khantha}}\ \emph {et~al.}(2004)\citenamefont
  {{Khantha}}, \citenamefont {{Cordero}}, \citenamefont {{Molina}},
  \citenamefont {{Alonso}},\ and\ \citenamefont
  {{Girifalco}}}]{2004PhRvB..70l5422K}%
  \BibitemOpen
  \bibfield  {author} {\bibinfo {author} {\bibfnamefont {M.}~\bibnamefont
  {{Khantha}}}, \bibinfo {author} {\bibfnamefont {N.~A.}\ \bibnamefont
  {{Cordero}}}, \bibinfo {author} {\bibfnamefont {L.~M.}\ \bibnamefont
  {{Molina}}}, \bibinfo {author} {\bibfnamefont {J.~A.}\ \bibnamefont
  {{Alonso}}}, \ and\ \bibinfo {author} {\bibfnamefont {L.~A.}\ \bibnamefont
  {{Girifalco}}},\ }\href {\doibase 10.1103/PhysRevB.70.125422} {\bibfield
  {journal} {\bibinfo  {journal} {\prb}\ }\textbf {\bibinfo {volume} {70}},\
  \bibinfo {pages} {125422} (\bibinfo {year} {2004})}\BibitemShut {NoStop}%
\bibitem [{\citenamefont {{Chuang}}\ \emph {et~al.}(2011)\citenamefont
  {{Chuang}}, \citenamefont {{Lin}}, \citenamefont {{Huang}}, \citenamefont
  {{Hsu}}, \citenamefont {{Kuo}}, \citenamefont {{Ozolins}},\ and\
  \citenamefont {{Yeh}}}]{2011Nanot..22A5704C}%
  \BibitemOpen
  \bibfield  {author} {\bibinfo {author} {\bibfnamefont {F.-C.}\ \bibnamefont
  {{Chuang}}}, \bibinfo {author} {\bibfnamefont {W.-H.}\ \bibnamefont {{Lin}}},
  \bibinfo {author} {\bibfnamefont {Z.-Q.}\ \bibnamefont {{Huang}}}, \bibinfo
  {author} {\bibfnamefont {C.-H.}\ \bibnamefont {{Hsu}}}, \bibinfo {author}
  {\bibfnamefont {C.-C.}\ \bibnamefont {{Kuo}}}, \bibinfo {author}
  {\bibfnamefont {V.}~\bibnamefont {{Ozolins}}}, \ and\ \bibinfo {author}
  {\bibfnamefont {V.}~\bibnamefont {{Yeh}}},\ }\href {\doibase
  10.1088/0957-4484/22/27/275704} {\bibfield  {journal} {\bibinfo  {journal}
  {Nanotechnology}\ }\textbf {\bibinfo {volume} {22}},\ \bibinfo {pages}
  {275704} (\bibinfo {year} {2011})}\BibitemShut {NoStop}%
\end{thebibliography}%


\end{document}